\documentclass{JINST}

\usepackage[pdftex]{graphicx}
   \DeclareGraphicsExtensions{.pdf,.jpeg,.png}
\usepackage{url}

\title{A novel 4D fast track finding system using precise space and time information of the
  hit}

\author{Nicola Neri$^a$\thanks{Corresponding
author.}, Marco Petruzzo$^b$\\
\llap{$^a$}INFN, Sezione di Milano\\
  Via Celoria 16, Milano, Italy\\
  E-mail: \email{nicola.neri@mi.infn.it}\\
  \llap{$^b$}INFN, Sezione di Milano and Universit\`a di Milano\\
    Via Celoria 16, Milano, Italy\\
}

\abstract{
  We propose a novel fast track finding system capable of reconstructing four
  dimensional particle trajectories in real time using precise space and
  time information of the hits. 
  Recent developments in silicon pixel detectors achieved $150~\textrm{ps}$  time resolution 
  and intense R\&D is in progress to improve the timing performance, aiming at $10~\textrm{ps}$.
  The use of the precise space and time information allows the suppression of background hits
  not compatible with the time of passage of the particle and the determination of its time evolution.
  The fast track finding device that we are proposing is based on a massively parallel algorithm
  implemented in commercial field-programmable gate array using a pipelined architecture. 
  We describe the algorithm and its implementation for a tracking system prototype based on 8 planes of silicon sensors
  used as a case study. According to simulations the suppression of noise hits is effective in 
  reducing fake track combinations and improving real-time track reconstruction in presence of background
  hits. The system provides offline-like tracks with sub-microsecond latency and it is capable to determine
  the time of the track with picosecond resolution assuming 10 ps resolution for the hits.
}


\begin{document}

\section{Introduction}

Recent developments in silicon pixel detectors achieved very good space and time resolutions, {\it i.e.}
the GigaTracker \cite{Fiorini:2013xya} of the NA62 experiment at the Super Proton Synchrotron at CERN
achieved resolutions of $\sigma_x = 100~\mu \textrm{m}$ and $\sigma_t = 150~\textrm{ps}$ for space and time, respectively.
Studies on silicon 3D pixel sensors showed that a time resolution of $180~\textrm{ps}$ can be
obtained~\cite{Parker:2011zz} .
In addition intense R\&D for fast timing detectors is in progress and sensors with improved time resolution
are foreseen to be available for applications in experiments in the next years.
A few examples are represented by microchannel plates PMTs~\cite{Frisch_TIPP2014} aiming at a resolution of about
$1~\textrm{ps}$ and $\sigma_x = 1~\textrm{mm}$ and ultra fast silicon detectors~\cite{Cartiglia:2013haa}
aiming to achieve time resolution of 10 ps and space resolution of 10 $\mu$m.

The precise time information of the hits can be used to
suppress noise hits not compatible with the expected time of passage  of the particles.
This feature would allow to use another dimension for discriminating signal and background events,
 particularly important for future experiments at high luminosities.
 The high luminosity phase of LHC (HL-LHC) is set to start after 2025 and aims to increase the total number of
collisions by a factor 10 with a foreseen instantaneous luminosity of $L=5\cdot10^{34}~\textrm{cm}^{-2}\textrm{s}^{-1}$.
At the HL-LHC an integrated luminosity of $250~\textrm{fb}^{-1}\textrm{year}^{-1}$ will be reached
and a large number of proton-proton interactions will be produced at each bunch crossing, on average
about $140$ primary interactions. This effect is indicated as pile-up.
As a consequence detectors will be exposed to a higher level of radiation with huge data rates
produced. Present trigger solutions adopted by LHC experiments would not be capable to reduce data rates
without a significant loss of information and physics potential.
A viable solution to overcome part of these problems is to include the tracking information in the early
stages of the trigger chain.
At high luminosity the pile-up becomes very important and association of tracks to different primary vertexes
is more difficult.
The use of fast-timing detectors can help in identifying tracks from different proton-proton interactions mitigating
the pile-up effects.

Fast track finding systems with low level trigger capabilities have been proved to be crucial for high energy physics
experiments at hadron colliders, {\it e.g.} the silicon vertex trigger of CDF experiment~\cite{Ashmanskas:1999ze},
and similar solutions are currently adopted by the ATLAS experiment~\cite{FTK}.
Major upgrades are foreseen for the HL-LHC phase of the CMS experiment, especially in the design of the new silicon tracker
 and detector readout, in order to use the tracking information for first level trigger decisions~\cite{CMS-phase2}.
\par
 In this work we propose a novel 4D real time tracking system using the precise space and time information
of the hits that can be used for the low level trigger and we show how it improves the background rejection
and the tracking performance in presence of noise hits.

\section{4D artificial retina algorithm}
The artificial retina algorithm for fast track reconstruction is described in Ref.~\cite{Ristori:2000vg} and
 it is inspired by neurobiology.
Cellular units distributed in the space of track parameters are tuned to identify specific charged particle
trajectories and provide response on how well a set of hits is matched to specific track hypotheses.

 



\begin{figure}[!h]
\centering
\includegraphics[width=0.95\textwidth]{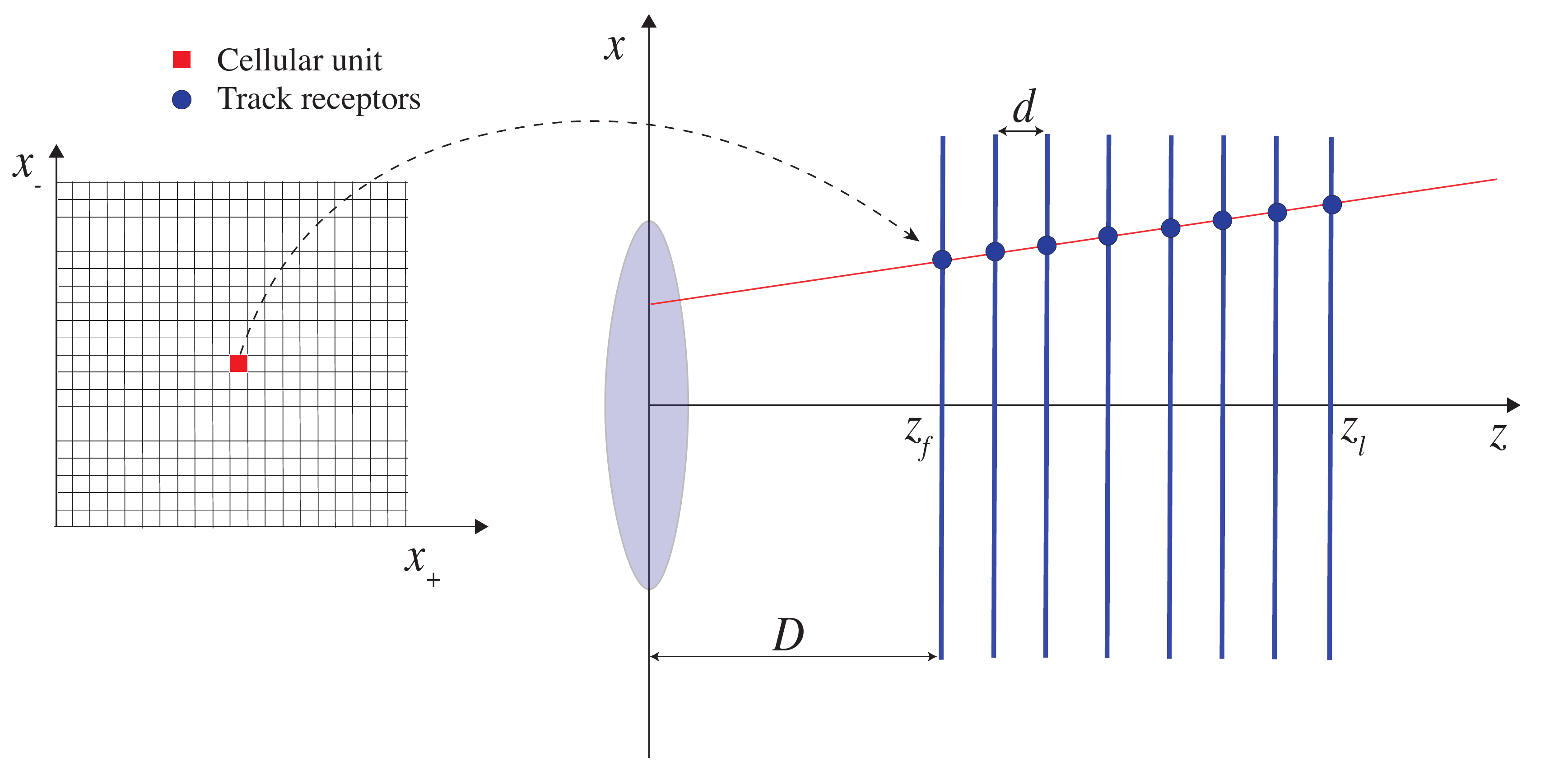}
\caption{
  Layout of the tracking detector used for simulations:
  the distance between the planes is $d=4$ cm and the first plane is placed
  at a distance $D=10$ cm from the interaction region, distributed along the $x$ axis.
}
\label{fig:layout}
\end{figure}

For simplicity sake let's consider a two dimensional tracking system based on single-sided silicon strip detectors.
However, the artificial retina algorithm can be generalized to work for three dimensional tracks as well~\cite{LHCb:TPU,Abba:2014gia},
using both pixel and strip detectors.
Let's define ($x_f, z_f$) and ($x_l, z_l$) the coordinates of the intersections of the tracks in the first and last
layer, respectively, of the 8-layer tracking detector shown in Fig.~\ref{fig:layout}.
We define the constant terms $z_\pm = (z_f \pm z_l)/2$ that depend on
 the geometry of the detector, and the track parameters $x_{\pm} =(x_f\pm x_l)/2$ are used to describe the 
 equation of a 2D track,
\begin{equation}
\label{eq:2D_track}
  x(z) = x_+ + x_- (z-z_+)/z_-.
\end{equation}
The time of the track is defined as the time of the particle when crossing the $x$ axis,
assuming the particle travels at the speed of light \textit{c}. This is a very good assumption for
 charged particles produced at the LHC, at center-of-mass energy of 14 TeV.

The space of track parameters ($x_+,x_-$) can be divided in a grid of cellular units. 
Each engine corresponds to a specific point ($i, j$) in the grid of track parameter space and is associated
with the track parameters ($x_{i+}, x_{j-}$).
A set of corresponding receptors is identified with the intercepts of the track in the detector layers,
as shown in Fig.~\ref{fig:layout}.
The expected time of the hit at layer $k$ is evaluated according to
\begin{equation}
\label{eq:time_track}
t_{ij}(z_k) = t_\textrm{trk}+z_k/c \sqrt{1+x^2_{j-}/z^2_{-}}
\end{equation}
where $t_\textrm{trk}$ is the time of the track a $z=0$.
The distance of the track receptor to the measured $k$-th hits
\begin{equation}
\label{eq:distance_track}
s_{ijk}=|x_k-x_{i+}-x_{j-}(z_k-z_+)/z_-|,
\end{equation}
is fed to the engine which calculates the weight according to a Gaussian field response, $\exp(-s^2_{ijk}/2\sigma^2)$,
where $\sigma$ is a parameter to be adjusted for optimal response.
The receptors also provide an additional Gaussian response for the time of the hit, $\exp(-t^2_{ijk}/2\sigma_t^2)$,
where $t_{ijk}=|t_k-t_{ij}(z_k)|$ is the difference between the time of the $k$-th measured hit and its expected value,
and $\sigma_t$ is the width of the field response. 
The engine weight function, $W_{ij}$, is defined as the sum over non negligible contributions 
\begin{equation}
\label{eq:weightsum}
W_{ij} = \sum_k W_{ijk} 
\end{equation}
where $W_{ijk}$ is the contribution from the $k$-th hit, defined as
\begin{eqnarray}
\label{eq:weight}
&W_{ijk} = & \exp{\left(-\frac{s_{ijk}^2}{2\sigma^2}\right)} \exp{\left(-\frac{t_{ijk}^2}{2\sigma_t^2}\right)} \textrm{\quad if } s_{ijk}<2\sigma ,  \\
&W_{ijk} = & 0 \hspace{2cm} \textrm{\hspace{2.6cm} if }  s_{ijk}>2\sigma. \nonumber 
\end{eqnarray}
The weight is evaluated for three different $t_\textrm{trk}$ hypotheses,
$t_\textrm{trk} = t_0 - \Delta T,~ t_0,~ t_0 + \Delta T$ where $t_0$ is the nominal bunch crossing time determined by
 the accelerator reference clock, and $\Delta T$ is proportional to the duration of the bunch crossing.
\begin{figure}[!h]
\centering
\includegraphics[width=0.45\textwidth]{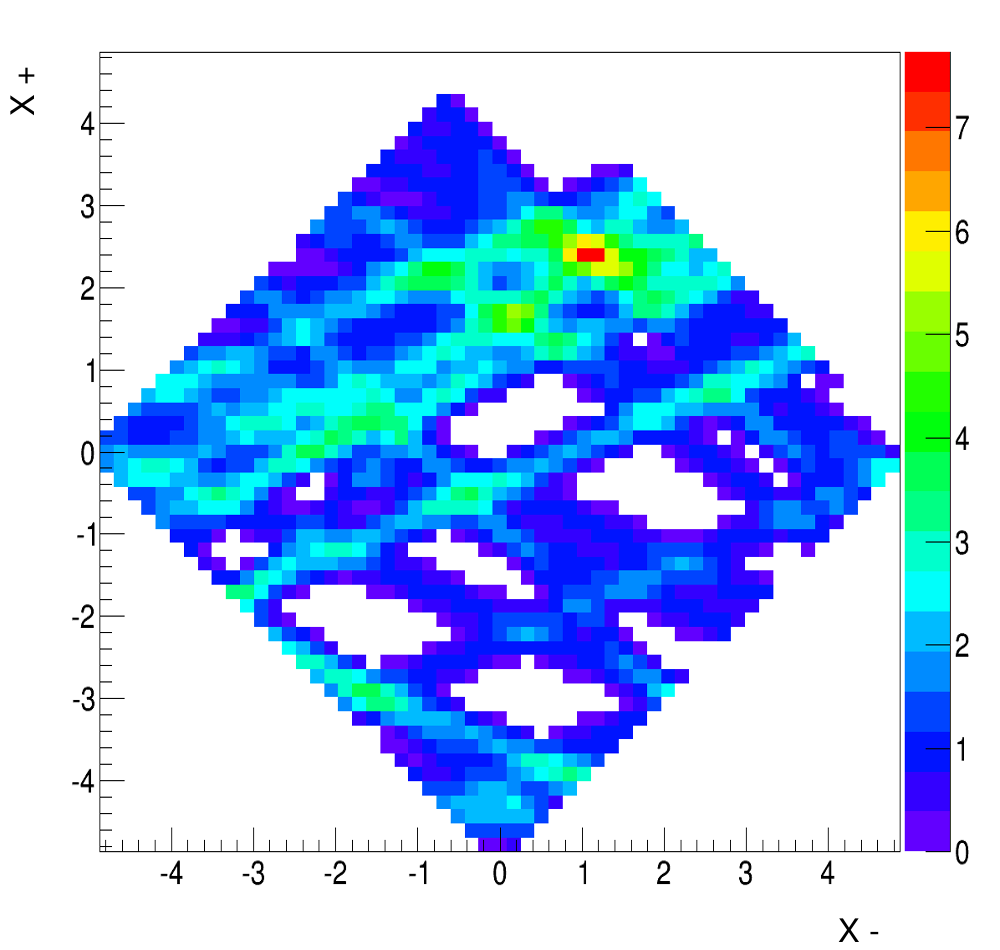}
\includegraphics[width=0.45\textwidth]{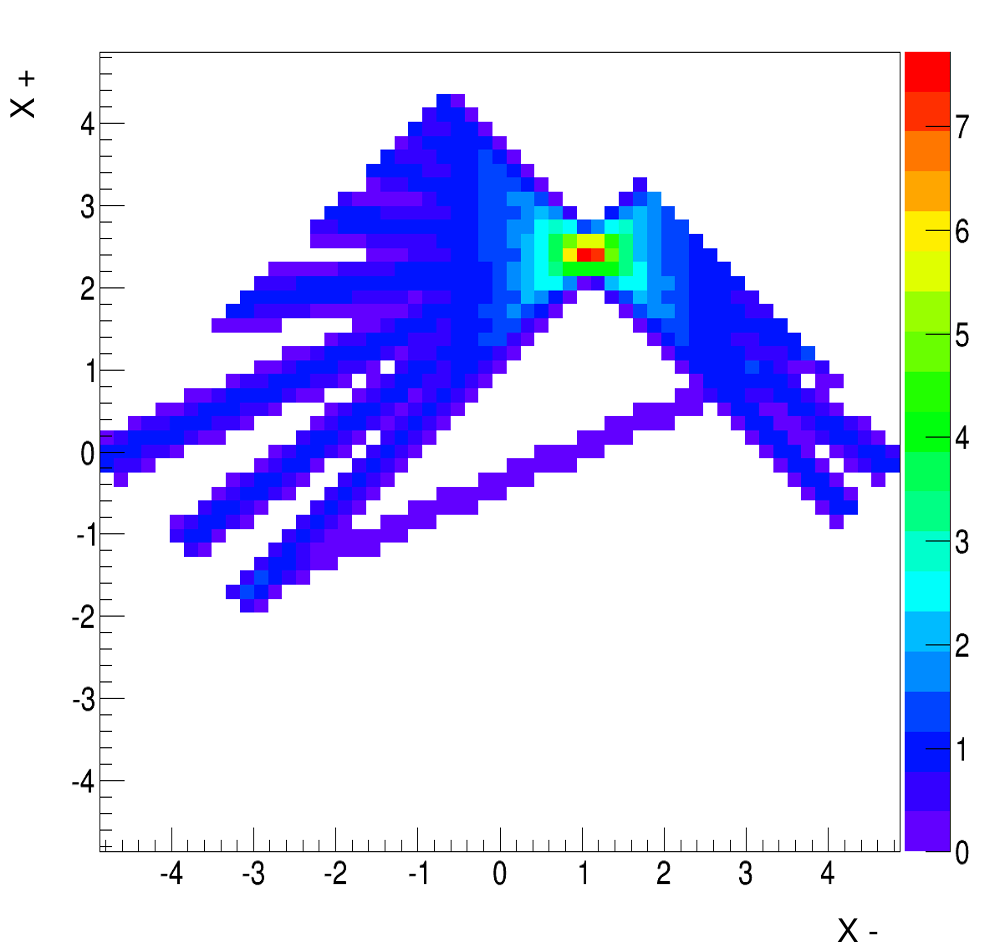}
\caption{
The retina response represented by the weight distribution, $W_{ij}$, in the grid of track parameters
for an event with a real track plus background hits (5\% detector occupancy) with no time information (left)
and using time information of the hit (right) with 10 ps resolution.
Tracks are identified by the local maxima of the weight distribution.
}
\label{fig:track_time}
\end{figure}

The evaluation of the weight function is performed in parallel for the appropriate engines and a track is identified by the local maximum at time $t_\textrm{trk}=t_0$.
The reconstructed space and time parameters of the track $(x_{+},x_{-},t_\textrm{trk})_{\textrm{rec}}$ are obtained by interpolations of the weight values adjacent to the maximum along the $x_+$, $x_-$ and $t$ axes. 

In particular the track parameters are determined by means of a Gaussian interpolation, defined as 
\begin{eqnarray}
  x_{+,\textrm{rec}} = x_{i+} + \frac{\Delta_{x_+}}{2}\frac{\ln( W_{i-1 j}/W_{ij})-\ln( W_{i+1 j}/W_{ij})}{\ln( W_{i-1 j}/W_{ij})+\ln( W_{i+1 j}/W_{ij})}, \\
  x_{-,\textrm{rec}} = x_{j-} + \frac{\Delta_{x_-}}{2}\frac{\ln( W_{i j-1}/W_{ij})-\ln( W_{i j+1}/W_{ij})}{\ln( W_{i j-1}/W_{ij})+\ln( W_{i j+1}/W_{ij})}
\end{eqnarray}
where $\Delta_{x_\pm}$ is the granularity of the grid of track parameters.
Comparable results are obtained determining the parameters with a weighted average, 
\begin{equation}
x_{+,\textrm{rec}} = \frac{\sum_i W_{ij} x_{i+} }{\sum_i W_{ij} }, \quad \quad x_{-,\textrm{rec}} = \frac{\sum_j W_{ij} x_{j-} }{\sum_j W_{ij}}.
\end{equation}

Similarly the time of the track can be evaluated by using a Gaussian interpolation along the time
of the maximal weight value at grid position $(i,j)$, 
\begin{eqnarray}
  t_{\textrm{trk},\textrm{rec}} = t_0 + \frac{\Delta T}{2}\frac{\ln( W_{i j t_-}/W_{ijt_0})-\ln( W_{i j t_+}/W_{ijt_0})}{\ln( W_{i j t_-}/W_{ijt_0})+\ln( W_{ij t_+}/W_{ijt_0})}
\end{eqnarray}
where $t_\pm= t_0\pm\Delta T$.

The contributions from hits not compatible with the time of the track are suppressed when including the time information in the retina algorithm.
In Fig. \ref{fig:track_time}, the weight distribution in the grid of track parameters is shown for an event
with a real track in presence of background hits with no time information (left) and using time information
of the hit (right).
The plots correspond to a simulation with detector occupancy of 5\% and resolution on the time of the hit of 10 ps.
The suppression of out-of-time hits helps reducing local maxima in the weight distribution, shown in Fig. \ref{fig:track_time}, that could generate fake tracks identification while improving the track parameter resolution, as discussed in Sec.~\ref{sec:simulations}.


\section{4D tracking simulations}
\label{sec:simulations}

At the LHC the typical r.m.s. of the longitudinal size of the colliding proton bunches is 7.5 cm and the r.m.s.
of the transverse size is few $\mu\textrm{m}$.
In this scenario the proton-proton interactions are distributed in a region of about 10 cm, with an average of 1.4
interactions/mm.
The primary interactions are distributed in time in an interval of about 330 ps around the nominal event time $t_0$,
determined by the accelerator reference clock while noise hits are uniformly distributed in time.
The maximum value of the weight along $t$ is expected in the central bin at $t_\textrm{trk}= t_0$, and the weight values for $t_\textrm{trk}= t_0 \pm \Delta T$ are used for determining the time of the track by interpolation.
In this study we set $\Delta T = 400$ ps.

Simulations of the response of the 4D retina algorithm for a telescope made of 8 single-sided silicon
strip detectors positioned along the $z$ axis have been performed.
The strip pitch of the sensors is 180 $\mu$m and the time resolution of the hits has been set to 10 ps.
The distance between the detector layers is 4 cm and tracks are originated from an interaction region
along the $x$ axis with Gaussian profile and positioned at a distance of 10 cm from the first plane,
as represented in Fig.~\ref{fig:layout}.
Events with one track within the telescope acceptance have been generated and the artificial retina
response has been simulated using a grid of about 20,000 cellular units uniformly distributed
in the track parameter region with a granularity of 0.47 mm for both $x_{-}$ and $x_{+}$.

%
%
%
Simulations of events with different level of detector occupancy (noise hits) have been performed
with and without the time information of the hits to evaluate the effect on the track parameter
resolutions.
The obtained resolutions for the $x_-$ and $ x_+$ track parameters  are $\sigma_{x_-}=24~\mu\textrm{m}$
and  $\sigma_{x_+}=15~\mu\textrm{m}$, respectively, for events with no noise hits;
the time of the track is determined with a precision of 3.5 ps.
This is compatible with $\sigma_t= 10/\sqrt{N} ~\textrm{ps}$ where $N=8$ is the number of measured hits.
\begin{figure}[!h]
\centering
\includegraphics[width=0.70\textwidth]{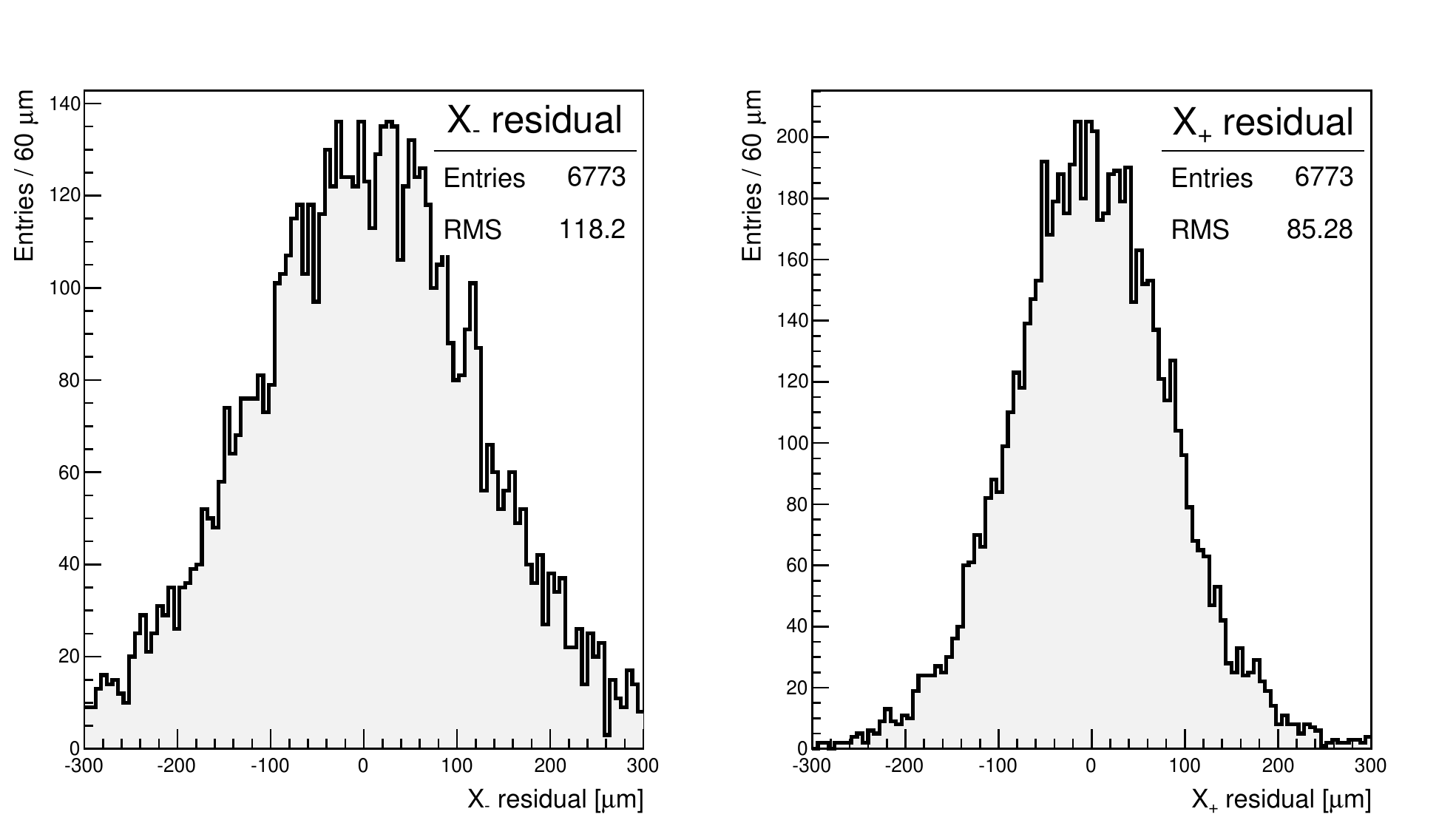}
\includegraphics[width=0.70\textwidth]{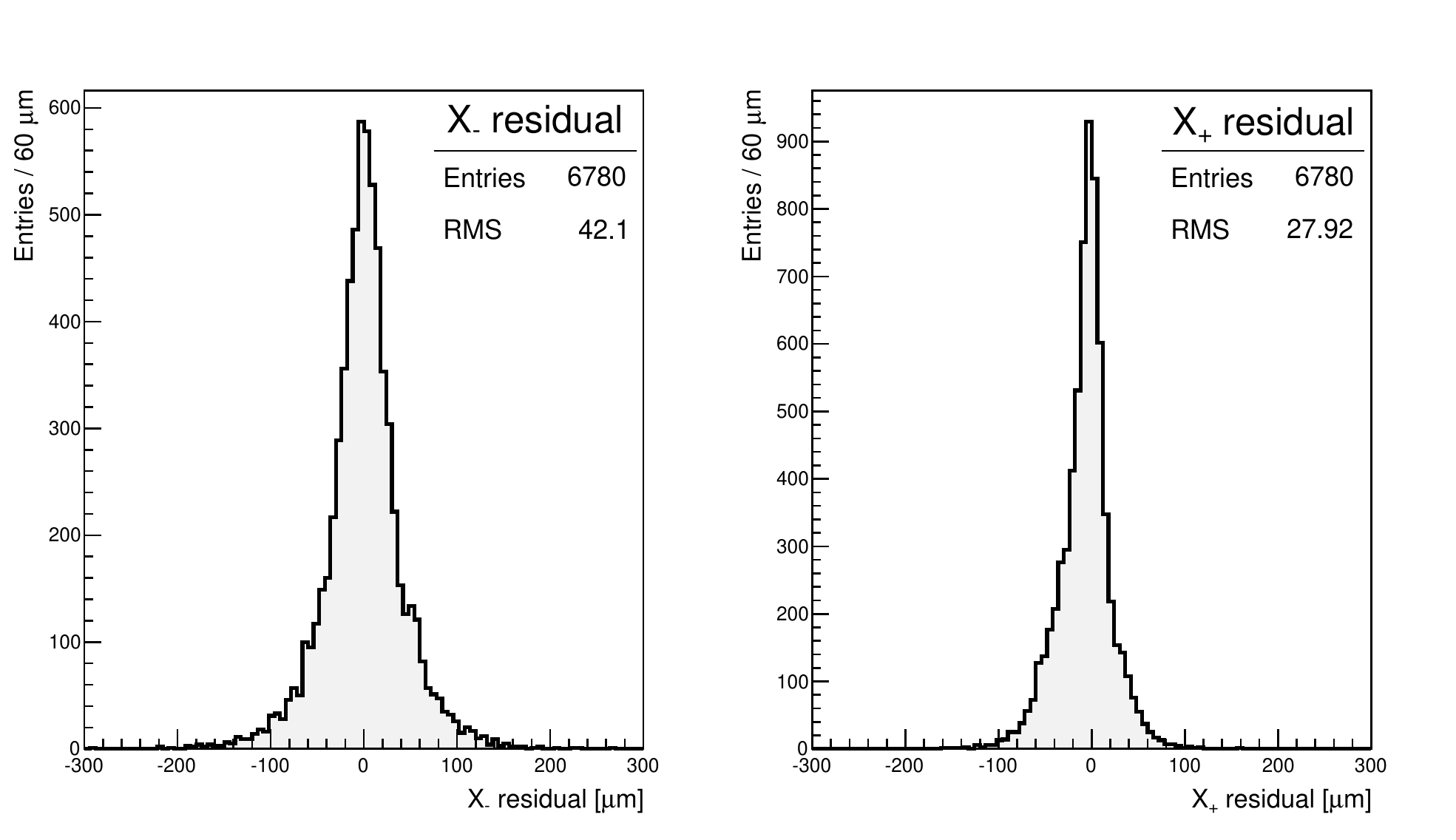}
\caption{
  Residual distribution for reconstructed minus generated $x_{-}$ and  $x_{+}$ track parameters 
  for simulated events with 5\% detector occupancy
  with no time information (top) and using time information of the hits with 10 ps resolution (bottom).
The resolution improves when using the time information and changes from $\sigma_{x_-}=118~\mu \textrm{m}$ and $\sigma_{x_+}=85~\mu \textrm{m}$ to $\sigma_{x_-}=42~\mu \textrm{m}$ and $\sigma_{x_+}=28~\mu \textrm{m}$.
}
\label{fig:resX}
\end{figure}

\begin{figure}[!h]
\centering
\includegraphics[width=0.45\textwidth]{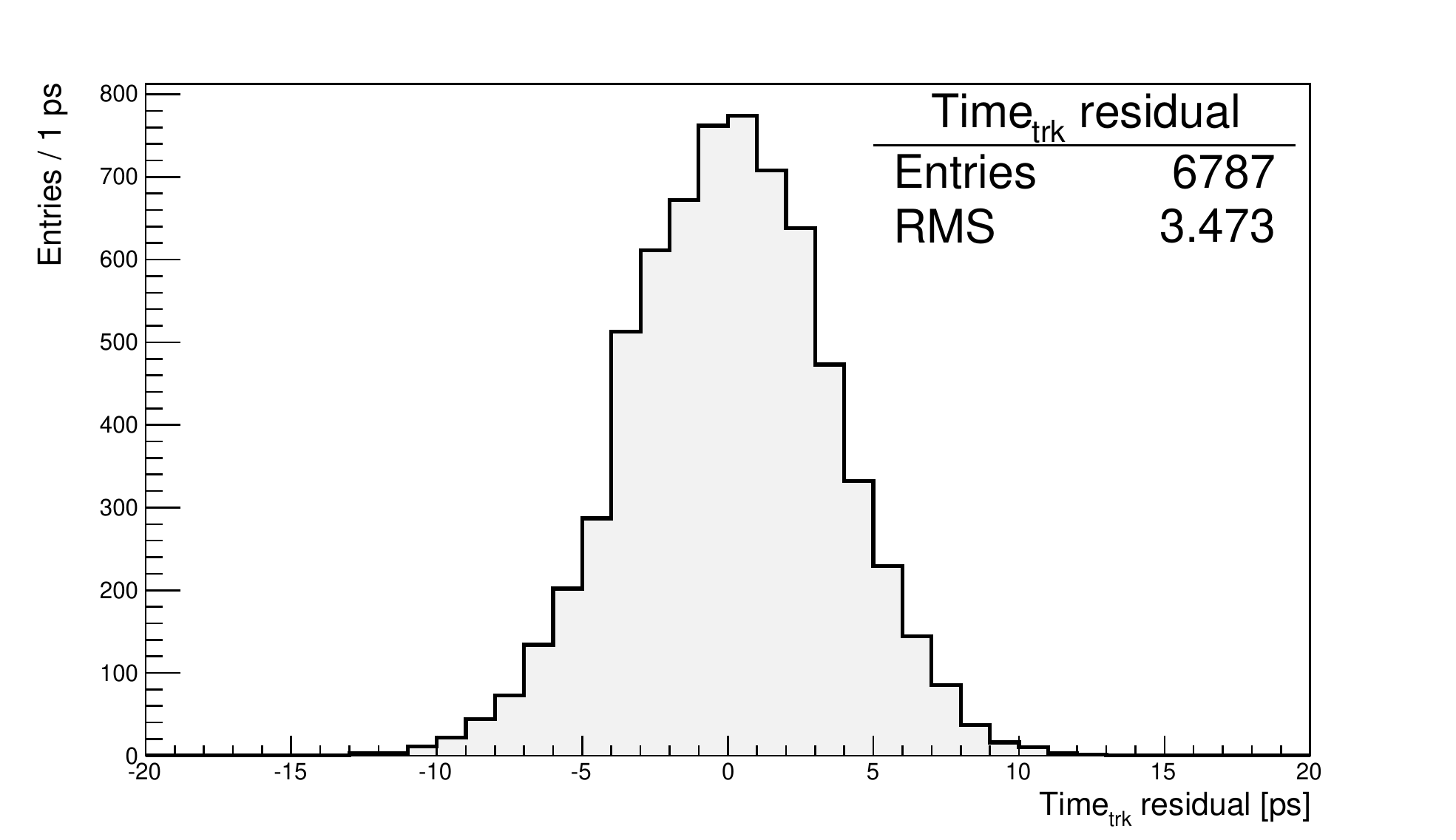}
\includegraphics[width=0.45\textwidth]{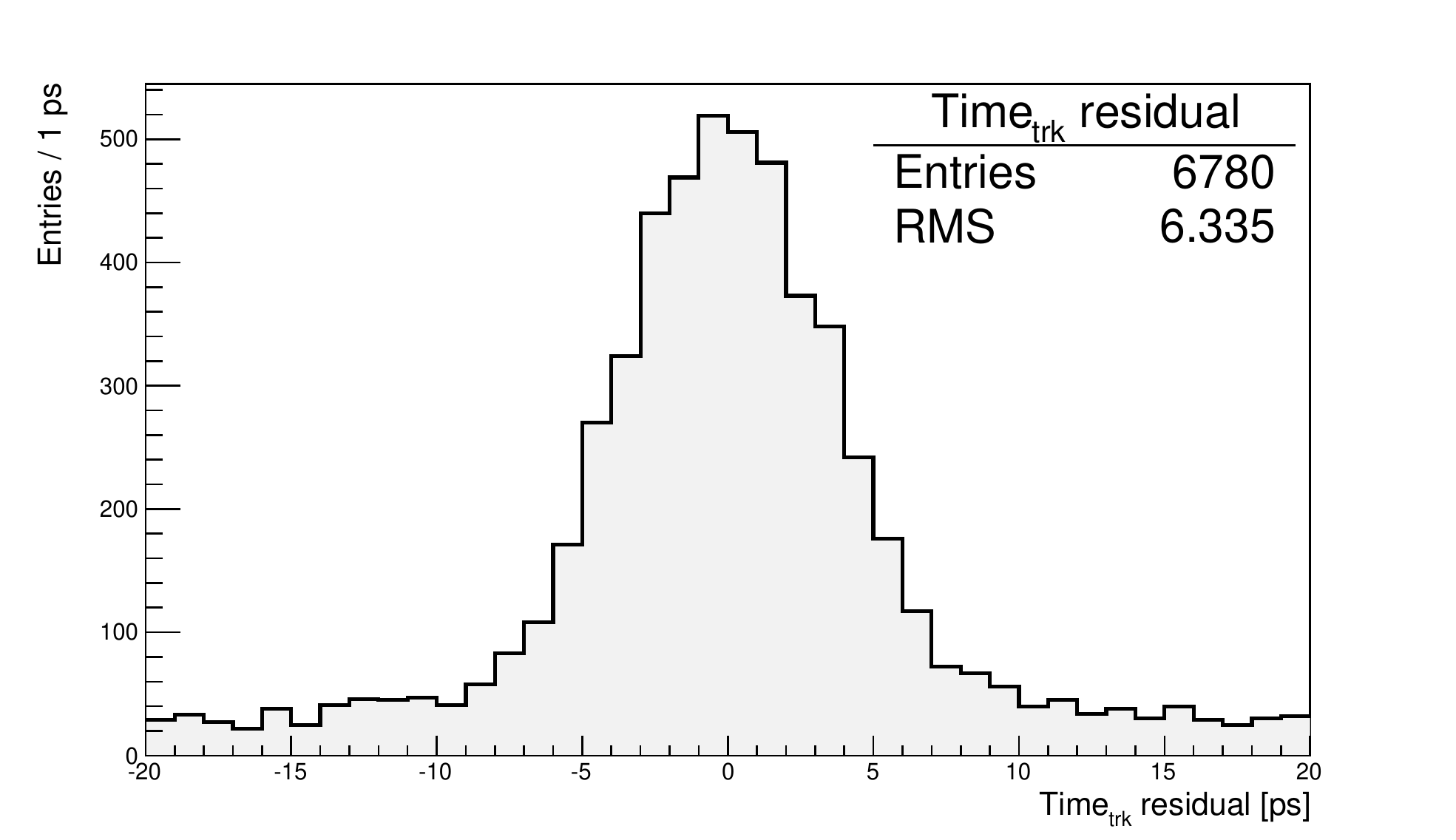}
\caption{
  Residual distribution for reconstructed minus generated time of the track for
  simulated events with no noise (left) and with noise and 5\% detector occupancy (right).
  The obtained resolution is $\sigma_t = 3.5$~ps with no noise (left) and 6.3 ps with noise hits (right);
  the time information of the hits has 10 ps precision and eight measured hits per track.}
\label{fig:resT_noise}
\end{figure}

When increasing the detector occupancy, background hits affect the quality of track reconstruction.
The worsening of the resolution is in general due to the contamination of the weight function near the
local maximum corresponding to an identified track.
This effect can be reduced by increasing the granularity of the grid of cellular units and, as we demonstrate here,
mitigating the effect of  background out-of-time hits by using the time information in the fast tracking algorithm.
In Fig. \ref{fig:resX} are shown the residual distributions of generated minus
 reconstructed $x_{-}$ and $x_{+}$ track parameters  for events with 5\% detector occupancy.
The resolutions obtained without using the time information are $\sigma_{x_-}=118~\mu\textrm{m}$ and $\sigma_{x_+}=85~\mu\textrm{m}$.
The performance decreases with respect to the case of events without noise hits.
When including the time information of the hit the tracking performance improves 
and the resolution on track parameters become $\sigma_{x_-}=42~\mu \textrm{m}$
and $\sigma_{x_+}=28~\mu \textrm{m}$.
The results of the simulations of the 4D artificial retina algorithm are summarised in Table~\ref{tab:simulation_results}
and compared with resolutions obtained with offline track reconstruction by means of a simple $\chi^2$ fit.
The residual distribution for the reconstructed minus the generated time of the track is shown in
Fig. \ref{fig:resT_noise} for simulated events with noise hits.
The obtained resolution is $\sigma_t = 6.3 ~\textrm{ps}$, to be compared with nominal value of $3.5$ ps for
the case with no noise.
The system provides a good determination of the time of the track even in presence of noise hits.
The precise information on the time of the track can be used at later stages of event reconstruction
to distinguish among particles coming from vertexes close in space but originated from proton-proton interactions
occurring at different times.

\begin{table}[htb]
  \caption{Resolutions on track parameters for simulated events with no noise hits, and with noise hits and detector occupancy of 5\%. Results of the 4D artificial retina algorithm are compared
    with resolutions obtained with offline track reconstruction.}
\label{tab:simulation_results}
\centering
\begin{tabular}{|l|c|c|c|c|}
\hline
Configuration & time info & $\sigma_{x_-}$ ($\mu$m) & $\sigma_{x_+}$ ($\mu$m) & $\sigma_{t_\textrm{trk}}$ (ps) \\
\hline
no noise, offline & no & 23  & 15 &  - \\
no noise, retina & no & 24 & 15 &  - \\
no noise, retina & yes & 24 & 15 & 3.5\\
with noise 5\% occ., offline & no & 26 & 18 & - \\
with noise 5\% occ., retina & no & 118 & 85 & - \\
with noise 5\% occ., retina & yes & 42 & 28 & 6.3 \\
\hline
\end{tabular}
\end{table}

\section{Implementation of the 4D artificial retina algorithm in commercial FPGAs}
The 4D artificial retina algorithm has been implemented on custom electronic boards developed in collaboration
with Nuclear Instruments~\cite{ref:NI} based on Xilinx Kintex7 field-programmable gate array (FPGA).
Using 512 cellular units for the grid of track parameters with a granularity of 3.3 mm
the response of the algorithm was simulated at a clock frequency of 200 MHz.
The choice of 512 engines is motivated by an existing implementation of the artificial retina algorithm
with no time information~\cite{NSS15:Neri} and found useful for comparing the utilization of resources
in the two different cases.

\begin{figure}[!h]
\centering
\includegraphics[width=0.90\textwidth]{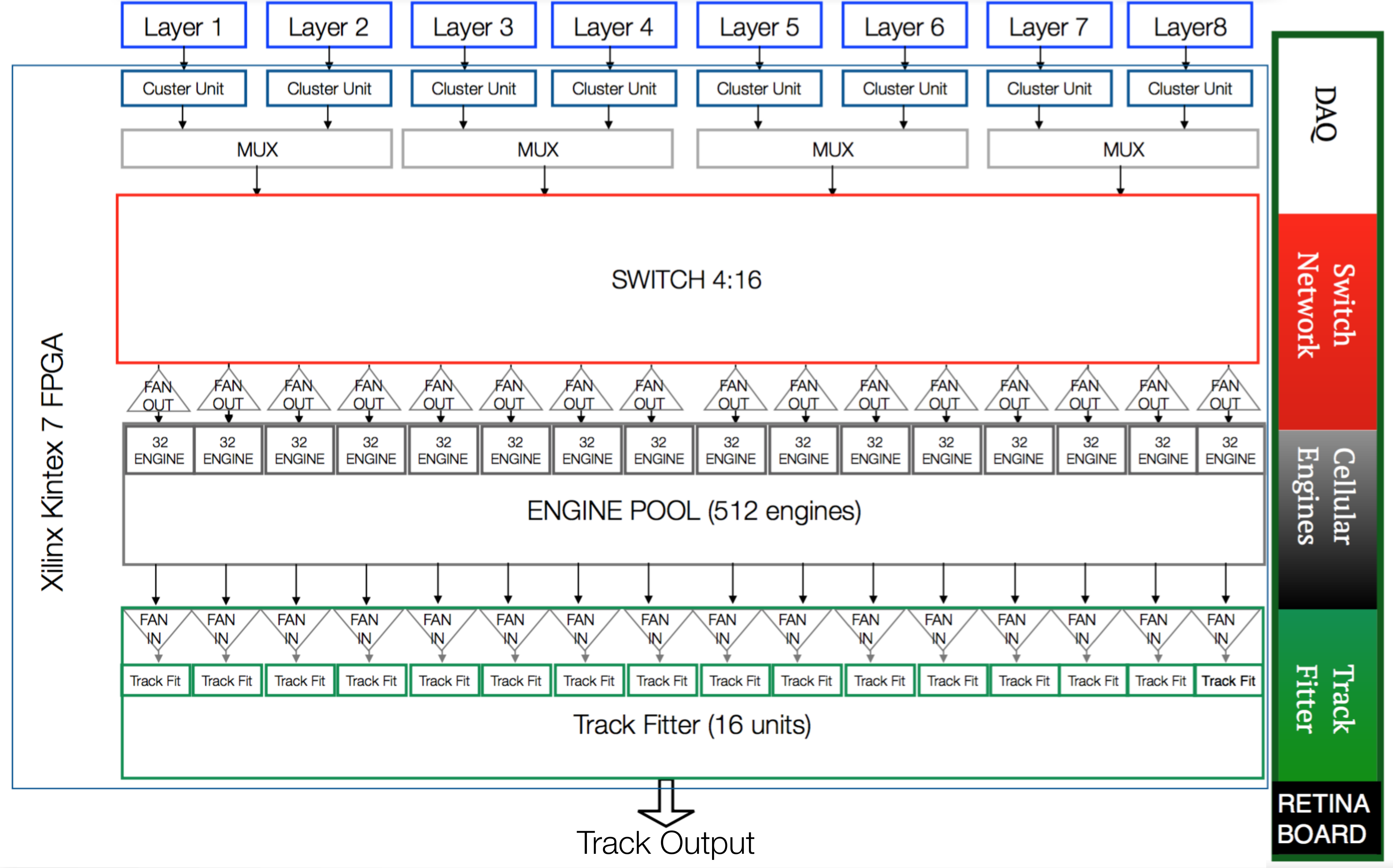}
\caption{Architecture of the real-time tracking system based on the artificial retina algorithm as
  implemented in a custom electronic board based on commercial FPGA.}
\label{fig:retina_architecture}
\end{figure}

The fast track finding architecture based on the artificial retina algorithm is composed by three different modules:
the switch module that delivers the hits from the detector layers to appropriate processing units,
a pool of engines for the calculation of the artificial retina response, and the track fitter module for the calculation
of the track parameters by interpolation of the weight values.  The latency of the response is estimated to be
less than 100 time units: about 14 time units for the switch, 15 time units for the engine, 30 time units for the track
fitter. This corresponds to a latency shorter than a microsecond at 200 MHz clock frequency,
{\it i.e.} 1 time unit = 5 ns. The architecture of the system is shown in Fig.~\ref{fig:retina_architecture}.

\begin{figure}[!h]
\centering
\includegraphics[width=1.0\textwidth]{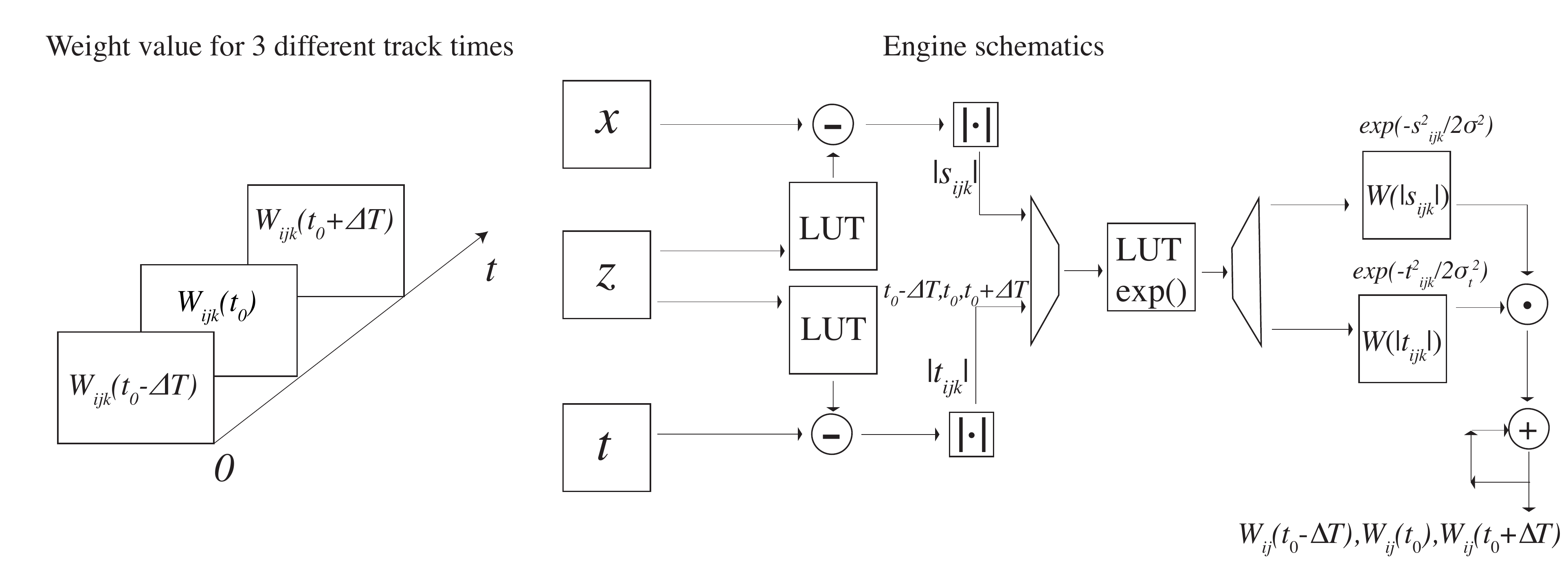}
\caption{The weight value for each cellular unit is calculated for three different track times $t_0-\Delta T, t_0$, and $t_0+\Delta T$ (left). Schematics of the engine processing unit for the calculation of weight (right).
}
\label{fig:engine_time}
\end{figure}

For the implementation of the 4D artificial retina algorithm, minor changes have been made to the switch and the 
track fitter logic modules with respect to the artificial retina algorithm implementation using spatial information only.
The switch delivers in parallel the information of the hits from the data acquisition to the appropriate engines
and the correct path for the hits is evaluated according to information stored in look-up tables (LUTs) .
In particular the data routing path depends only on the space information of the hit
and no change is required to the switch module.
For the case of the 4D artificial retina algorithm a modification to the architecture was necessary to the engine logic unit,
 shown in Fig.~\ref{fig:engine_time}.
The engine receives the hits with the information of the $z$ coordinate of the detector layer,
the position of the measured $x$ cluster position and its time $t$.
The values $|s_{ijk}|$ and $|t_{ijk}|$ are evaluated according to two different LUTs whose values depend on the
$z$ coordinate only. The exponentiation is performed in the FPGA using a LUT in order to save resources and
reduce the latency of the response.
The algorithm needs to evaluate a total of three weight values corresponding to the different track time hypotheses: $t_\textrm{trk}=t_0-\Delta T$, $t_0$, $t_0+\Delta T$.
For each incoming $k$-th hit the term $\exp(-s^2_{ijk}/2\sigma^2)$ is evaluated and then multiplied
for three different time factors, $\exp(-t^2_{ijk}/2\sigma^2_t)$.
The same LUT is shared for the exponentiations.
The four inputs are serialized and the outputs are deserialized allowing to instantiate only one LUT per engine.
An hold signal is propagated from the engine back to the switch module for three clock cycles, until the engine is able to accept a new hit.
The multiplications between space and time Gaussian responses are implemented using DSP (digital signal processing) blocks.
In the track fitter the $x_-$ and $x_+$ track parameters are evaluated, as for the case of the artificial retina algorithm,
by interpolation of the weight values of the cells adjacent to the local maximum ~\cite{NSS15:Neri}.
These modifications to the architecture of the system require a modest increase of FPGA resources of about 10\%.
\section{Conclusions and perspectives}
Existing silicon pixel detectors achieved 150 ps time resolution and strong effort is ongoing to improve the time
precision at the level of 10 ps. The use of precise space and time information for fast track finding systems
would allow to use an additional dimension for discriminating signal and background events.
A novel track finding algorithm using precise time and space information of the hit is proposed here.
The 4D artificial retina algorithm allows the precise determination of the time
of the track at the origin and a strong suppression of the contributions from background hits out of time. 
According to simulations the improvement in tracking performance is sizeable in presence of noise hits.

The algorithm is based on a massively parallel calculation and it is suitable for implementation in FPGAs 
with a pipelined architecture.
The algorithm has been implemented in a Xilinx Kintex 7 FPGA with a modest increase of the resources, about 10\%, with
respect to the case where no time information was considered.
These results are encouraging and represent a first step towards the development of a 4D fast track finding system
 for applications in future experiments at high luminosity.

\acknowledgments

This work is supported by Istituto Nazionale di Fisica Nucleare, Italy. The authors would like to thank Andrea Abba
for useful discussions.

\end{document}